\documentclass[12pt,thmsa]{article}
\usepackage{amssymb}

\usepackage{sw20lart}

\input{tcilatex}
\begin{document}

\author{E. S. Corchero \and Departamento de F\'{i}sica. Universidad de Cantabria.
Santander. Spain}
\title{Gravitational vacuum polarization as an alternative to dark matter }
\maketitle

\begin{abstract}
It is assumed that the quantum vacuum may be studied as consisting of two
contributions, with positive and negative energy respectively, which
interact but slightly and may be displaced from each other. Then it is
proposed that dark matter may be just an increase of the quantum vacuum
energy, with respect to the normal dark energy level, induced by the
gravitational field of galaxies or clusters. A simple model is worked out
able to reproduce astronomical observations.
\end{abstract}

An outstanding open problem in present day cosmology is the nature of ``dark
energy'' and ``dark matter'', which toghether contribute more than 95\% of
the total mass-energy of the universe. The dark energy corresponds to a mass
density 
\begin{equation}
\rho _{DE}\sim 10^{-26}kg/m^{3},  \label{00}
\end{equation}
and the dark matter to about one third this quantity on the average\cite
{Sahni}. A common hypothesis is to identify the dark energy with the
cosmological constant introduced by Einstein in 1917. An alternative
hypothesis, equivalent in practice, is to postulate a vacuum stress-energy
tensor, $T^{\mu \nu }$, of the form 
\begin{equation}
T^{\mu \nu }=\rho g^{\mu \nu },  \label{0}
\end{equation}
$g^{\mu \nu }$ being the metric tensor, which amounts to assuming an
equation of state for the vacuum of the form 
\begin{equation}
p=-\rho <0,\;\rho =T_{0}^{0},\;p=-T_{1}^{1}=-T_{2}^{2}=-T_{3}^{3}.
\label{0a}
\end{equation}

If the dark energy is due to the quantum vacuum, it is difficult to
understand why its density is not either strictly zero or a density at the
Planck scale, that is about $10^{123}$ times the observed value, eq.$\left( 
\ref{00}\right) .$ We might assume that the vacuum contains both positive
and negative energy contributions which do not cancel completely each other,
but then an extremely fine tuning is requiered in order to get the observed
value. In the present paper I propose that the dark energy derives from the
fact that the positive energy of long wavelength components of the vacuum
fields is not completely cancelled by the other field components whilst its
positive pressure is more than compensated by the negative pressure of other
field components. This hypothesis reduces substantially the fine tuning
needed and, in addition, leads naturally to the hypothesis that the vacuum
may be ``gravitationally polarized''. That is, vacuum components
contributing positive (negative) energy will tend to go to regions with more
(less) negative gravitational potential. The purpose of the present paper is
to show that this assumption may allow interpreting ``dark matter'' as a
gravitationally polarized ``dark energy''.

In more detail the argument is as follows. We should assume that all
short-wavelength components of the quantum vacuum are strongly coupled to
each other, as shown by the coupling of the corresponding vaccum
fluctuations. For instance, virtual photons with Compton wavelength $(\sim
10^{-12}m)$ may produce virtual electron-positron pairs and viceversa. This
suggests that short-wavelengths of the different vacuum fields could not be
separated from each other and their joint contribution to the vacuum energy,
either positive or negative, should be not too big in absolute value. In
contrast, vacuum field components with long wavelength may be almost
uncoupled to those with short wavelength, but either completely cancel the
short-wavelength contribution or give a very small, but positive, total
vacuum energy. This hypothesis offers the possibility that the total vacuum
energy and pressure, in the absence of gravitational field, fulfils the dark
energy eqs.$\left( \ref{00}\right) $ to $\left( \ref{0a}\right) $ without
the need of an extremely fine tuning.

Now it is natural to believe that the long wavelengths part of the vacuum
fields may propagate with some independence from the short wavelength
components. In order to get a simple model, I shall assume that there is a
cut-of, $\lambda _{c},$ such that field components with longer wavelength
are \textit{completely decoupled} from the remaining vacuum fields. The
model so derived is clearly too crude and a more plausible model should
include some interaction between the positive energy and the negative energy
contributions, thus preventing a too big vacuum polarization. This
possibility will be considered elesewhere. Here I shall pursue with the
development of the simple model by fixing the cut-off, $\lambda _{c}$. It is
obvious that it should lie somewhere between the Compton and macroscopic
wavelengths, but it is difficult to make a precise estimate. I shall choose $%
\lambda _{c}\sim 1\mu m,$ at about the wavelength of visible light, which
will lead to good agreement with known dark matter properties (see below).
This value corresponds to a particle mass 
\[
m\sim hc/\lambda _{c}\sim 1eV/c^{2}, 
\]
not far from neutrino masses. Thus wavelengths greater than $1\mu m$ are
negligible for vacuum fields other than the electromagnetic one.

The mass density associated to the electromagnetic zero-point field (energy $%
1/2h\nu $ per normal mode) cut-off at $\lambda _{c}$ is positive and has the
value
\begin{equation}
\rho _{0}=\frac{1}{c^{2}}\int_{0}^{c/\lambda _{c}}(\frac{1}{2}h\nu )\frac{%
8\pi \nu ^{2}}{c^{3}}d\nu =\frac{\pi h}{c\lambda _{c}^{4}}\sim 4\times
10^{-16}kg/m^{3}.  \label{1b}
\end{equation}
The essential hypothesis of the model here proposed is that the vacuum
contribution due to long wavelengths, with density eq.$\left( \ref{1b}%
\right) ,$ and the remaining vacuum contributions, with negative density,
behave as \textit{two non-interacting fluids}. In particular in a static
space-time, the only case to be studied here, each part should be in
hydrostatic equilibrium in the gravitational field.

For simplicity I shall consider a static space-time with spherical symmetry,
where I should study the hydrostatic equilibrium of three non-interacting
fluids, ruled by 
\begin{equation}
\frac{dp_{j}}{dr}=-\Phi ^{\prime }\left( \rho _{j}+p_{j}\right) ,\;j=1,2,3,
\label{1}
\end{equation}
where $j=1$ corresponds to the long wavelenghts vacuum contribution, $j=2$
to the remaining vacuum fields and $j=3$ to the baryonic matter. (Units $c\
=G=1$ will be used throughout this paper.) The gravitational potential, $%
\Phi ,$ is given by 
\begin{equation}
\Phi ^{\prime }\equiv \frac{d\Phi }{dr}=\frac{m+4\pi r^{3}p}{r^{2}-2mr}%
,\;m(r)=\int_{0}^{r}4\pi \rho \left( r\right) r^{2}dr,  \label{1a}
\end{equation}
where $\rho =$ $\rho _{1}+\rho _{2}+\rho _{3}$ is the total mass density and 
$p=p_{1}+p_{2}+p_{3}$ is the total pressure at $r$ (all measured in the
local frame). It may be realized that the sum in $j$ of the three eqs.$%
\left( \ref{1}\right) ,$ combined with eq.$\left( \ref{1a}\right) $ gives
the standard Tolman-Oppenheimer-Volkoff equation of general relativity,
namely 
\begin{equation}
\frac{dp}{dr}=-\frac{d\Phi }{dr}\left( \rho +p\right) \equiv -\frac{m+4\pi
r^{3}p}{r^{2}-2mr}\left( \rho +p\right) .  \label{1c}
\end{equation}

It is easy to relate the density and pressure of each component of the
vacuum with the potential $\Phi $ if we know the corresponding equation of
state. For the long-wavelength electromagnetic zero-point field I must
assume the radiation equation of state, that is 
\begin{equation}
p_{1}=\frac{1}{3}\rho _{1},  \label{2}
\end{equation}
which, inserted in the first eq.$\left( \ref{1}\right) ,$ leads to

\begin{equation}
\frac{1}{3}\frac{d\rho _{1}}{dr}=-\Phi ^{\prime }\frac{4}{3}\rho
_{1}\Rightarrow \rho _{1}=\rho _{10}\exp \left( -4\Phi \right) ,  \label{3}
\end{equation}
$\rho _{10}$ being the contribution to mass density at infinity (where $\Phi
\rightarrow 0$)$.$ As said above the contribution of the remaining vacuum
fields to both the energy density and the pressure should be negative, in
order to cancel almost completely the positive contribution of the
long-wavelength (electromagnetic) field. I shall assume that the second
contribution, seen as a fluid, is more rigid that the first one so that the
pressure as a function of the energy density is more steep than eq.$\left( 
\ref{2}\right) .$ As a simple equation of state which allows analytical
solutions I propose 
\begin{equation}
p_{2}=\gamma \frac{(\rho _{2})^{2}}{\rho _{20}},\;\rho _{20}<0,  \label{4}
\end{equation}
where $\gamma $ is a dimensionless parameter and $\rho _{20}$ is the
negative contribution to the mass density at infinity. Putting this into the
first eq.$\left( \ref{1}\right) $ I obtain 
\begin{equation}
\gamma \frac{1}{\rho _{20}}\frac{d(\rho _{2})^{2}}{dr}=-\Phi ^{\prime
}\left[ \rho _{2}+\gamma \frac{(\rho _{2})^{2}}{\rho _{20}}\right]
\Rightarrow \rho _{2}=\frac{\rho _{20}}{\gamma }\left[ \left( 1+\gamma
\right) \exp \left( -\frac{1}{2}\Phi \right) -1\right] .  \label{5}
\end{equation}
The sum of the two contributions will give the total energy density and
pressure of the vacuum, namely 
\begin{equation}
\rho =\rho _{10}\exp \left( -4\Phi \right) +\frac{\rho _{20}}{\gamma }\left[
\left( 1+\gamma \right) \exp \left( -\frac{1}{2}\Phi \right) -1\right] ,
\label{6}
\end{equation}
\begin{equation}
p=\frac{1}{3}\rho _{10}\exp \left( -4\Phi \right) +\frac{\rho _{20}}{\gamma }%
\left[ \left( 1+\gamma \right) \exp (-\frac{1}{2}\Phi )-1\right] ^{2}.
\label{7}
\end{equation}
Now it is trivial to find values of $\rho _{10},\;\rho _{20}$ and $\gamma $
such that, when $\Phi =0,$ we get the dark energy values, that is $\rho
=-p=\rho _{DE}$ with $\rho _{DE}$ given by eq.$\left( \ref{00}\right) .$ I
get 
\begin{equation}
\rho _{10}=(K+1)\,\rho _{DE},\;\rho _{20}=-K\,\rho _{DE}\text{ with }K\equiv 
\frac{4}{3\gamma -1}>>1.  \label{7a}
\end{equation}
Assuming $\rho _{10}\sim -\rho _{20}\sim \rho _{0}$ as given in eq.$\left( 
\ref{1b}\right) $ I obtain $K\sim 4\times 10^{10}$, a very high value which
means that the density and pressure of the two assumed vacuum components
cancel quite accurately.

Eqs.$\left( \ref{6}\right) $ and $\left( \ref{7}\right) $ may be simplified
taking into account that, in the cases of interest for us, the potential is
very weak, that is $\left| \Phi \right| <<1$. To second order in the
potential the equations become 
\begin{equation}
\rho =\rho _{0}\left[ \exp \left( -4\Phi \right) +3-4\exp \left( -\frac{1}{2}%
\Phi \right) \right] \simeq -2\rho _{0}\Phi +\frac{15}{2}\rho _{0}\Phi
^{2}+\rho _{DE},  \label{8}
\end{equation}
\begin{equation}
p=\frac{1}{3}\rho _{0}\exp \left( -4\Phi \right) -3\rho _{0}\left[ \frac{4}{3%
}\exp (-\frac{1}{2}\Phi )-1\right] ^{2}-\rho _{DE}\simeq \rho _{0}\Phi
^{2}-\rho _{DE}.  \label{9}
\end{equation}
Typically $\Phi \sim -10^{-6}$ in galaxies so that, identifying $\rho _{0}$
with the value eq.$\left( \ref{1b}\right) $ I obtain $-2\rho _{0}\Phi \sim
10^{-21}kg/m^{3}>>\rho _{DE}.$ In contrast the dark matter pressure
predicted by the model is of the same order as the dark energy pressure, $%
-\rho _{DE}$, but has opposite sign. The comparison with observations is as
follows. Recent measurements in the Milky Way\cite{Xue} give a mass $%
M(<60kpc)=(4.0\pm 0.7)\times 10^{11}M_{\circledcirc }$ so that the potential
is $\Phi (60kpc)\sim -M/R=-3.3\times 10^{-7}$, consistent with the above
estimate for $\Phi $ within the galaxy. On the other hand it is known\cite
{Berg} that the local density, near the solar system, is $%
0.2-0.5Gev/cm^{3}\sim \left( 4-9\right) \times 10^{-22}kg/m^{3},$ in
agreement with our model estimate.

Eq.$\left( \ref{8}\right) $ allows writing the ``vacuum polarization law''
in the simple form 
\begin{equation}
\nabla \rho \simeq -2\rho _{0}\nabla \Phi =2\rho _{0}\mathbf{g,}  \label{9a}
\end{equation}
$\mathbf{g}$ being the gravitational field. But I stress that this relation
should have limited validity. Indeed, in strong fields the vacuum
polarization would suffer some saturation so that $\left| \nabla \rho
\right| $ is smaller than predicted by eq.$\left( \ref{9a}\right) .$ On the
other hand for very small potential, $\Phi ,$ eqs.$\left( \ref{8}\right) $
and $\left( \ref{9}\right) $ might be modified because, being small
differences of big quantities, they would be rather sensitive to any change
in the model.

We see that the mass density, eq.$\left( \ref{8}\right) ,$ and pressure, eq.$%
\left( \ref{9}\right) ,$ due to the polarized vacuum energy mimic a fluid of
cold matter with equation of state 
\begin{equation}
p_{DM}=\frac{1}{4\rho _{0}}\rho _{DM}^{2}<<\rho _{DM},  \label{10}
\end{equation}
where $\rho _{DM}=\rho -\rho _{DE}$ is the density above the dark energy
level, and similarly for the pressure. Thus in the study of dark matter we
might use eq.$\left( \ref{10}\right) $ as the equation of state of a
hypothetical fluid. For instance if eq.$\left( \ref{10}\right) $ is put in
the hydrostatic equilibrium equation (see eq.$\left( \ref{1}\right) )$ I get
eqs.$\left( \ref{8}\right) $ and $\left( \ref{9}\right) $ to leading order
in $\Phi $, that is 
\begin{equation}
\rho _{DM}\simeq -2\rho _{0}\Phi ,\;p_{DM}\simeq \rho _{0}\Phi ^{2}.
\label{10a}
\end{equation}
Of course eqs.$\left( \ref{8}\right) $ to $\left( \ref{10}\right) $ are
specific for our simple model, in particular they rest upon eq.$\left( \ref
{4}\right) .$ However the qualitative behaviour of the polarized vacuum
energy mimicking cold matter is likely valid for any reasonable model, e. g.
with equation of state different from eq.$\left( \ref{4}\right) $.

In the following I will study the dark matter problem in a galaxy assumed
spherical. The study of more realistic galaxies and clusters would be
similar although more involved. For this study I may neglect the dark energy
contribution, which is very small whenever the dark matter is relevant. In
order to get the spatial distribution of dark matter I shall solve eq.$%
\left( \ref{1a}\right) .$ For this purpose the Newtonian approximation is
good enough, that is

\begin{equation}
\Phi ^{\prime }\simeq \frac{m}{r^{2}}\Rightarrow \frac{d}{dr}\left(
r^{2}\Phi ^{\prime }\right) =\frac{dm}{dr}=4\pi r^{2}\left( \rho +\rho
_{B}\right) =-8\pi r^{2}\rho _{0}\Phi +4\pi r^{2}\rho _{B},  \label{11}
\end{equation}
where $\rho _{B}\left( r\right) $ is the baryonic mass density of the galaxy
and I have approximated $\rho $ by the first term of the right hand side of
eq.$\left( \ref{8}\right) .$ In the external region, where baryonic matter
is negligible, the solution of eq.$\left( \ref{11}\right) $ is 
\begin{equation}
\Phi =-\frac{C}{r}\sin \left( \mu r+\delta \right) \Rightarrow \rho =\frac{%
2C\rho _{0}}{r}\sin \left( \mu r+\delta \right) ,  \label{12a}
\end{equation}
where 
\begin{equation}
\;\mu =\sqrt{8\pi \rho _{0}G/c^{2}},\;\mu ^{-1}\equiv R_{DM}\sim 3\times
10^{20}m\sim 10kpc.  \label{12}
\end{equation}
$R_{DM}$ is the typical radius of the dark matter distribution predicted by
the model and I have used for $\rho _{0}$ the value $\left( \ref{1b}\right) $%
. The solution in the internal region cannot be found without knowing the
distribution of baryonic mater. As a simple model I shall consider a
distribution consisting of a constant density, $\rho _{B},$ within a sphere
of radius $R$ , zero outside. Then the regular solution of eq.$\left( \ref
{11}\right) $ for $r<R$ is 
\begin{equation}
\Phi =\frac{\rho _{B}}{2\rho _{0}}-\frac{A}{r}\sin \left( \mu r\right)
\Rightarrow \rho =\frac{2A\rho _{0}}{r}\sin \left( \mu r\right) -\rho _{B}.
\label{16}
\end{equation}

The condition that both $\Phi $ and $\Phi ^{\prime }$ are continuous at $r=R$
leads to 
\begin{eqnarray}
A\sin \left( \mu R\right) -C\sin \left( \mu R+\delta \right)  &=&\frac{\rho
_{B}}{2\rho _{0}}R=\frac{4\pi \rho _{B}R}{\mu ^{2}},  \nonumber \\
A\cos \left( \mu R\right) -C\cos \left( \mu R+\delta \right)  &=&\frac{\rho
_{B}}{2\rho _{0}\mu }=\frac{4\pi \rho _{B}}{\mu ^{3}},  \label{17}
\end{eqnarray}
where eq.$\left( \ref{12}\right) $ has been taken into account. These
equations allow getting two of the parameters $A,C$ and $\delta $ in terms
of one of them. Thus I will write the dark matter distribution in terms of $%
\delta $ as follows. Multiplying the first eq.$\left( \ref{17}\right) $
times $cos\left( \mu R\right) $ and the second one times $sin\left( \mu
R\right) $ and subtracting I get 
\begin{equation}
C\sin \delta =\frac{4\pi R}{\mu ^{2}}\rho _{B}\left[ \frac{\sin \left( \mu
R\right) }{\mu R}-\cos \left( \mu R\right) \right] \simeq \frac{4}{3}\pi
R^{3}\rho _{B}=M_{B},  \label{17a}
\end{equation}
$M_{B}$ being the total baryonic mass of the galaxy. In the second eq.$%
\left( \ref{17a}\right) $ I have taken into account that the typical radius
of dark matter in galaxies is much larger than the radius of baryonic matter
, i. e. $\mu R=R_{baryonic}/R_{DM}<<1$. Putting eq.$\left( \ref{17a}\right) $
into the first eq.$\left( \ref{17}\right) $ I obtain 
\begin{equation}
A=M_{B}\left( \frac{1}{\tan \delta }+\frac{3}{2\mu R}+\frac{3}{\mu ^{3}R^{3}}%
\right) ,  \label{17c}
\end{equation}
where I have again neglected terms of order $\mu R.$ Using the expressions
obtained for $A$ and $C$ in eqs.$\left( \ref{12a}\right) $ and $\left( \ref
{16}\right) ,$ respectively, I get the potential, to order $\mu R,$ and the
dark matter density in terms of the parameter $\delta ,$ namely 
\begin{eqnarray}
\Phi  &\simeq &\left( -\frac{3}{2}-\frac{\mu R}{\tan \delta }+\frac{r^{2}}{%
2R^{2}}\right) \frac{M_{B}}{R},  \nonumber \\
\;\rho  &\simeq &2\rho _{0}\left( \frac{3}{2}+\frac{\mu R}{\tan \delta }-%
\frac{r^{2}}{2R^{2}}\right) \frac{M_{B}}{R},\;r<R,  \label{17b} \\
\Phi  &\simeq &-\frac{M_{B}}{r\sin \delta }\sin \left( \mu r+\delta \right)
,\;\rho \simeq \frac{2M_{B}\rho _{0}}{r\sin \delta }\sin \left( \mu r+\delta
\right) ,\;r>R.  \label{17d}
\end{eqnarray}
Provided that $\pi /2>\delta >\mu R,$ as I shall assume (see next
paragraph), the latter expression shows that, beyond the typical galaxy
radius $R$, the dark matter density decreases as $1/r$ at the beginning and
more steeply later on, going to zero at $r=(\pi -\delta )/\mu .$ This
behaviour roughly agrees with observations.

\smallskip It remains $\delta $ as a free parameter which cannot be obtained
from our model. Thus the normalization (amplitude) of the dark matter
distribution is arbitrary and, furthermore, the model predicts bound states
of dark matter alone, without any amount of real (e. g. baryonic) matter.
This possibility seems unplausible. Consequently we should assume that such
bound states are not stable, but stability cannot be studied within our
time-independent model. Improvements of the model will be studied elsewhere,
but in the present paper I will simply fix the amplitude of the dark matter
distribution by comparison with observations. The total dark matter mass may
be obtained integrating the density eqs.$\left( \ref{17b}\right) $ and $%
\left( \ref{17d}\right) $ in the region where $\rho >0,$that is 
\begin{equation}
M_{DM}=\int 4\pi \rho r^{2}dr\sim \frac{M_{B}}{\sin \delta }\int_{\mu
R}^{\pi -\delta }x\sin \left( x+\delta \right) dx\sim M_{B}\left( \frac{\pi
-\delta }{\sin \delta }+1\right) ,  \label{20a}
\end{equation}
where the dark matter mass within the sphere of radius $R$ has been
neglected, it being of order $(\mu R)^{2}$ ($1/tan\delta $ is of order $\mu
R,$ see below). As the dark matter mass is typically more than 10 times the
baryonic mass in galaxies\cite{Berg}, the choice $sin\delta \lesssim 0.3$
seems appropriate.

For large $r$ the density eq.$\left( \ref{17d}\right) $ oscillates between
positive and negative values, which seems unphysical. We cannot use the
expedient of substituting $\rho =0$ for the predicted density $\left( \ref
{17d}\right) $ when $r$ is larger than the first zero of the density
function. Indeed this would contradict the relation eq.$\left( \ref{8}%
\right) $ between dark matter density and potential. Actually when $r$ is
large, and therefore $\left| \Phi \right| $ small, we should take into
account several corrections, e. g. deviations from sphericity, the influence
of neighbour galaxies, or the presence of some baryonic matter in the form
of gas. I shall study just one correction, namely the dark energy term $\rho
_{DE}$ which should be added to $\rho .$ In order to see whether this term
makes the density positive everywhere we should take into account that the
minimum of the density eq.$\left( \ref{17d}\right) $ will happen near $\mu
r\sim 3\pi /2$ with the value 
\[
\rho _{\min }\sim -\frac{2\rho _{0}\mu M_{B}}{\left( 3\pi /2-\delta \right)
\sin \delta }\sim -0.1\rho _{0}\mu R\frac{M_{B}}{R}\frac{1}{\sin \delta }.
\]
Thus with $\rho _{0}$ given by eq.$\left( \ref{1b}\right) $, $\mu R\sim
0.1,\sin \delta \sim 0.3,$ $M_{B}/R\sim 10^{-6}$ , I get $\rho _{\min }\sim
-10^{-23}kg/m^{3},$ far from being cancelled by the dark energy density, eq.$%
\left( \ref{00}\right) .$ Thus the predicted total density, $\rho +\rho
_{DE},$ is still negative in some regions, which shows that the model
requires some modifications. These would be most important for very small
densities where the cancellation between the $\rho _{1\text{ }}$and $\rho
_{2}$ is delicate (see eq.$\left( \ref{6}\right) ).$

A prediction of the model here proposed is the universal value of the dark
matter radius, $\mu ^{-1}$, estimated in about $10kpc$. Thus the extension
of the region occupied by dark matter is ruled by the parameter $\mu ,$ with
the result that it will be concentrated in the central region in clusters
(whose radius is larger than $\mu ^{-1})$ whilst it would extend well beyond
the region of baryonic matter in galaxies, which roughly agrees with
observations\cite{Sahni}. Of course this property is also true if dark
matter consists of a gas of particles with an equation of state like eq.$%
\left( \ref{10}\right) .$

\smallskip In summary I have shown that both dark matter and dark energy
might be interpreted as vacuum mass-energy provided that the vacuum may be
(slightly) polarized by the presence of gravity. The particular model here
proposed roughly agrees with observations. It is to be seen whether a more
sophisticated model may reproduce all the observed properties of dark matter.


\begin{thebibliography}{9}
\bibitem{Sahni}  Varun Sahni, Dark matter and Dark energy, Lect. Notes Phys.
653, 141-180 (2004). Astro-ph/0403324v3.

\bibitem{Xue}  X. X. Xue et al., e-print Archive astro-ph/0801.1232 (2008).

\bibitem{Berg}  Lars Bergstr\"{o}m, Rep. Prog. Phys. 63, 793-841 (2000).
\end{thebibliography}
\end{document}